\documentclass[pre,floats,superscriptaddress]{revtex4}

\usepackage{graphicx}
\usepackage{epstopdf}
\usepackage{amssymb}
\usepackage{amsmath,bm}
\usepackage{psfrag}
\usepackage{epsfig}
\usepackage{float}
\newcommand{\angstrom}{\mbox{\normalfont\AA}}

\begin{document}
\title{Morphologies, metastability and coarsening of quantum nanoislands on the surfaces of the annealed  Ag(110) and Pb(111) thin films}

\author{Donald L. Price}
\affiliation{Department of Mathematics, Western Kentucky University, Bowling Green, KY 42101, USA}
\author{Victor Henner}
\affiliation{Department of Theoretical Physics, Perm State University, Perm, 614990 Russia}
\affiliation{Department of Mathematics, Perm National Research Polytechnic University, Perm, 614990, Russia}
\author{Mikhail Khenner\footnote{Corresponding
author. E-mail: mikhail.khenner@wku.edu.}}
\affiliation{Department of Mathematics, Western Kentucky University, Bowling Green, KY 42101, USA}
\affiliation{Applied Physics Institute, Western Kentucky University, Bowling Green, KY 42101, USA}

\begin{abstract}

Morphological evolution of heteroepitaxial nanoislands toward equilibrium (coarsening) is computed using the detailed continuum model
that incorporates the quantum size effect.
Results reveal the metastability of the ``magic" heights, show the morphological transitions and the surface diffusion routes by which a 
quantum island 
reaches its stable height, and provide the coarsening laws for the island density and area, thus clarifying
the kinetic morphology pathways in the growth of an ultrathin metal films.

\textit{Keywords:}\ Heteroepitaxial ultra-thin metal films, two-step film growth, quantum size effect, ``magic" heights, metastability, coarsening, surface diffusion model.
\end{abstract}

\date{\today}
\maketitle


\section{Introduction}
\label{Intro}

Self-organized metal nanoisland arrays are important for the emerging technologies based on the nonlinear optics, 
plasmonics, photovoltaics, and photocatalysis. Such arrays hold large promise in the design of the spectrally selective absorbers \cite{Held},
the enhanced fluorescence and infrared spectroscopic devices \cite{Aslan,Jensen}, the polarizers and spectral filters \cite{Heger}, and the molecular detectors and biosensors \cite{Anker, Liao}.
They are also of interest for manufacture of the next-generation solar cells \cite{Santbergen}.

\begin{figure}[h]
\centering
\includegraphics[width=3.0in]{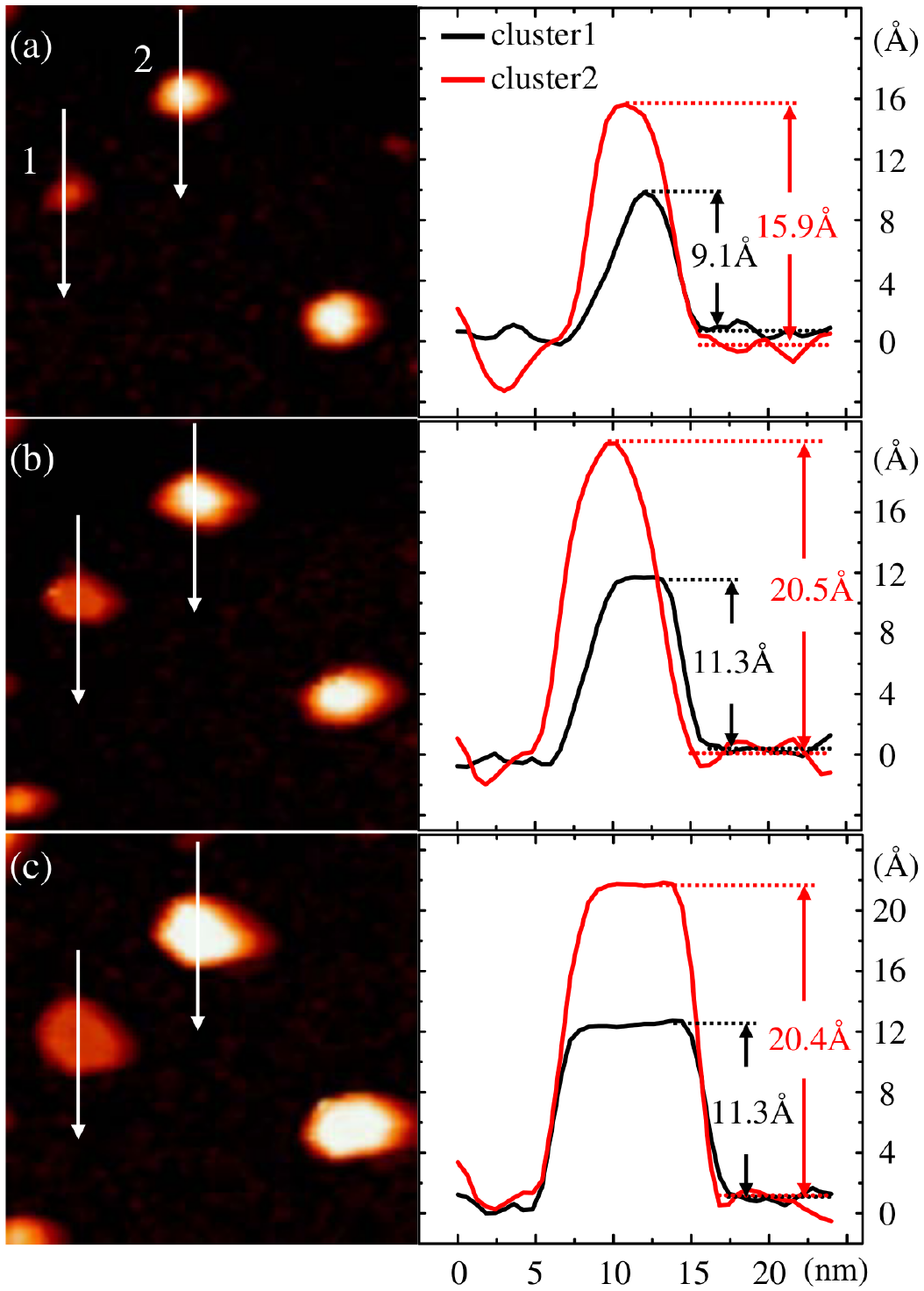}
\caption{(Color online.) (a-c): STM images of the growth of two 3D clusters. The line profiles to the right of (a-c)
represent the morphology evolution of the clusters 1 and 2 along the arrows. \copyright IOP Publishing. Reproduced from Ref. \cite{SCT} with permission. All rights reserved.
}
\label{Fig_Su_etal}
\end{figure}

Nanoislands with the flat (atomically smooth) top surfaces, steep edges, and strongly preferred heights in the interval 3-25 atomic monolayers (ML) can be formed on the surfaces of 
an ultrathin metal films by a two-step
heteroepitaxial growth technique, which consists of a low temperature ($<$140K) deposition, followed by a prolonged annealing at a moderate temperature (250-400K) 
\cite{Smith,OVM,DKGTH,Han1,SCT,Ozer,UFTE,Li,CHPPM,Gavioli,WMJFJJX,HHTWHT,YJEWWNZS,Hirayama}. Because of the complicated spectra of the preferred heights
such islands are often called ``magic". The two-step growth is called the ``electronic" or 
``quantum" growth, since the origin of the preferred island heights lies in the quantization of the electronic energy in the direction across the 
metal film \cite{Zhang,Ozer,Han1}. This quantization (quantum well states) is the result of the energy barriers at the film/substrate and film/vacuum interfaces, and it leads to oscillations of 
many physical properties, such as the surface energy, electrical resistivity, surface adhesion, thermal expansion coefficient, surface adsorption energy, work function, etc. 
with the film thickness. Precisely, the oscillations are the result of ``the systematic variation in the (electron)
density of states at the Fermi level due to its periodic crossing by the quantum well states created by the confinement of electrons" \cite{CHPPM}.
The total electronic energy falls in local minimums at specific film thicknesses \cite{Zhang}, thus, the film tends to align its local thickness to a specific one. 
The result typically is atomically flat film morphology that minimizes the electronic energy.
This so-called quantum size effect (QSE) also impacts heteroepitaxial growth of islands, often replacing the Stranski-Krastanow (SK)-type 
island morphology (the islands with a rounded tops and without a preferred heights) with the above-described ``quantum" morphology.
For example, for Ag(110) and Pb(111) ultra-thin films whose morphological evolution 
we model in this work, Table \ref{Tab2} shows the spectra of preferred (``magic") heights from several models and experiments, and Fig. \ref{Fig_Su_etal} makes clear the morphology difference
between the quantum and SK islands. 
At various temperatures and coverages, the morphology of 
an ultrathin metallic film grown by a two-step method may be a bi-continuous network \cite{WMJFJJX,Hirayama}, or it may be a large 
number of distinct islands \cite{SCT,Ozer,UFTE,Li,CHPPM,Gavioli}. In either case the heights of these features are at large selected from a set of the preferred heights 
particular to the material system. Detailed understanding of the kinetic pathways to form the ``magic"  island morphologies remains incomplete. 

In Refs. \cite{MSMSE_Khenner,JEM_Khenner} a basic surface diffusion equation-based continuum model was developed for studies of the evolving quantum island morphologies during 
annealing.
The model was applied to a generic film/substrate system featuring the \emph{ad hoc} simply-periodic surface energy oscillation (that is, without a beating pattern).
In this paper, that model is suitably modified and tuned for
computation of the nanoislands on Pb(111) and Ag(110) surfaces. Two-step island growth on these surfaces has been studied in the experiment \cite{OVM,DKGTH,Han1,
SCT,Ozer,UFTE,Li,CHPPM,WMJFJJX,HHTWHT,YJEWWNZS}. The substrates traditionally used in the experiments include 111 and 110-oriented Si, Cu, Ge, Fe, GaAs, and NiAl single-crystal films.   
Also, the theories of QSE for Pb(111) and Ag(110) ultrathin films were developed \cite{Han,OPL}, which allows us 
to construct the corresponding continuum models of the (oscillatory) surface energy that result in close match of the calculated and experiment sets of preferred heights.
The summary of the relevant data from the models and experiments is shown in Table \ref{Tab2}.

Modeling studies of the quantum nanoislands ensembles have been undertaken in Refs. \cite{Li,Kuntova}. 
Ref. \cite{Li} presents the rate equation-type model. Pb islands are taken as circular mesas with the radii $R_i$ and the heights $h_i$, and the evolution equation for 
the number of atoms in the $i$-th island is constructed and solved numerically. The exchange of atoms between the islands is through a dense wetting layer. The chemical
potential of this layer is phenomenologically introduced as a function of its coverage.
This model was applied to compute islands coarsening, assuming that the heights $h_i$ are the constant values from the initial condition 
(even when the $i$-th island species eventually disappears), all islands have the same constant circular mesa morphology, and there is no epitaxial stress influence on the
morphology evolution (despite that the lattice-mismatch stress in Pb/Si system is large). Computations in Ref. \cite{Li} show that the evolution of a fraction of area of 
islands with different heights closely matches the experiment.
In Ref. \cite{Kuntova} Monte Carlo simulations of the post-deposition evolution of stable Pb islands are presented. In this model the islands also have constant
circular mesa morphology. Island evolution is driven by the competition of the probabilities $\pi_1$ and $\pi_2$, where the former probability is one of an atom jump from a 
wetting layer onto island's sidewall (increasing the island width), and the
Arrhenius probability $\pi_2$ is one of an atom jump from a sidewall onto the top of the island (increasing the island height). 
These probabilities are chosen according to the complicated semi-empirical
rules which phenomenologically reflect the influence of epitaxial strain, with the values of some parameters that are known, at best, only to their order of magnitude.
Besides, since the island stability is assumed, the model cannot track the densities of metastable islands. Nonetheless, the computations in Ref. \cite{Kuntova}
found the regions of the parameter space that correspond to the islands of different heights.
Both models \cite{Li,Kuntova} were applied at a single
low temperature, thus the effects of the annealing were not studied.

Su et al. \cite{SCT} conducted the extensive quantum growth and annealing experiments with the ultrathin Pb films on Si(111). Besides the findings that are universally cited for 
quantum growth, 
such as the spectra of a ``magic" island heights, these authors notice a non-standard features, for instance, the co-existence of the round-top and flat-top islands and the transition 
from the round-top to flat-top  islands.  (Their terms for round-top and flat-top islands are three-dimensional (3D) clusters and 2D islands, respectively.) The latter feature implies that
3D clusters may be the seeds for growing 2D islands. In order to confirm this, Su et al. \emph{in situ} observe the growth process of an individual island with STM.
This is shown in Fig. \ref{Fig_Su_etal}, which is reproduced from their paper. As the deposited amount of Pb increases from (a) to (c), two round-top islands transform into the flat-top
islands, which then start growing laterally (2D growth). Following this, Su et al. show that growth of the initial nucleus is dominated by the SK growth mode, which causes it
to grow into a round-top island. The round-top to flat-top transition occurs once the round-top island aggregates enough atoms, which is necessary for QSE to 
become sufficiently strong that it can overcome the SK growth. Once an island acquires the flat top, its subsequent growth is in the 2D mode. They also demonstrate that in order 
for a 3D to 2D transition to occur, right before the transition the height of the round-top island must reach the ``magic" height, and then the flat-top island emerges with that height.
In other words, the $N$-layer flat-top island is transformed from the $N$-layer round-top island, provided that N belongs to the set of ``magic" values,
and thus the flat-top islands of each ``magic" height have a unique growth pathway. The 3D-to-2D transition probabilities are different for the round-top islands of different
``magic" heights (with the sixteen times difference between the most probable and the least probable, see their Fig. 4(e)), which may explain why a certain ``magic" height is
found more abundant in any growth experiment to-date \cite{SCT,Ozer,UFTE,Li,CHPPM,Gavioli}. 

%

The above summary of the existing models of the quantum nanoislands ensembles highlights their strengths and limitations.
The model in this paper, in which QSE and epitaxial stress are competing naturally and the surface morphology is dynamic, is better suited for understanding Su et al. \cite{SCT} 
experiments and other similar experiments \cite{CHPPM,WMJFJJX,YJEWWNZS,Hirayama}.
The surface atomic currents in this non-probabilistic model stem from a single self-consistent surface diffusion equation that does not differentiate 
between the different sections of the surface, allowing for the ``magic" island heights to emerge or disappear through the classical surface diffusion-driven dynamics \cite{Mullins}, 
provided only the initial non-planar surface morphology. 
We investigate the path of the 
nanoislands to the equilibrium by computing the morphological evolution of a set of the nanoislands that initially have the random sizes and positioning on the surface of a 
film. These initial islands are non-quantum, round-top SK islands without the preferred heights, as described in Sec. \ref{EvolveIslands}. 
Our computations span the full evolution from the different, random initial island distributions to the equilibrium state of a single island.
To facilitate this goal, 2D model was chosen for computational time savings. Despite the reduced dimensionality, our model successfully captures the transition to the electronic 
growth mode, where the islands grow laterally with a constant ``magic" height.
We find many
features of the island evolution that are in accord with Su et al. observations, and also go beyond that by (i) clarifying the metastability of various ``magic" heights, 
(ii) revealing very detailed morphological transitions, and (iii) computing the coarsening laws.

It must be noted that the continuum models based on the variants of the classical surface diffusion equation \cite{Mullins} (without the inclusion of the electronic effects) 
have been successfully applied to SK growth of the strained ultrathin solid films (thickness $7 - 20$ML) \cite{AFV,AF}.
Other examples of the continuum models that enable the computation of the dynamic morphologies of an ultrathin films can be found in Ref. \cite{MSMSE_Khenner}.
Recently, several sophisticated 2D and 3D continuum models of thicker film epitaxial growth and morphological evolution have been published \cite{SFA,SFAA,DKM,LD,WJBS,WS,SBVM}.





\section{Model of the morphological evolution of Pb and Ag quantum nanoislands}
\label{Formulation}


Our refined continuum model is based on the following 2D surface diffusion equation \cite{SDV,Chiu,GolovinPRB2004,Korzec,Korzec1,Golubovic,SGDNV,OL,SFAA,AFV,AF}: 
\begin{equation}
h_t=\sqrt{1+h_x^2}\ \mathcal{D}\frac{\partial^2}{\partial s^2}\left(\mu+\Sigma\right),
\label{MullinsEq}
\end{equation}
where the function $z=h(x;t)$ is the height, above the substrate, of the curve representing the film surface in 2D, $x$ is the coordinate along 
the substrate, $z$ is the coordinate normal to the substrate and into the film (with $z=0$ corresponding to the substrate), $s$ 
is the arclength along the curve, $\mathcal{D}=\Omega^2 D \nu/k T$ is the diffusion constant ($\Omega$ is the adatom volume, $D$ the adatom diffusivity, 
$\nu$ the surface density of the adatoms, $k T$ the Boltzmann's factor), $\Sigma$ is the elastic strain energy density of the film,
and 
\begin{equation}
\mu = \gamma(h) \kappa+n_3 \frac{d\gamma(h)}{dh}
\label{mu_kappa_mu_wet}
\end{equation}
is the chemical potential \cite{Chiu,GolovinPRB2004,Korzec,Korzec1,Golubovic,SFAA,AFV,AF}.
Here $\gamma(h)$ is the surface energy density, $\kappa$ is the surface curvature, and $n_3$ is the $z$-component of ${\bf n}$. 
The second term in Eq. (\ref{mu_kappa_mu_wet}) is the contribution of wetting.
The strain energy is assumed to result due to the film and substrate lattice mismatch. In Appendix, we briefly discuss the 
epitaxial stress model that we employ. The intrinsic surface stress is not considered, since it is negligible in comparison with the lattice-mismatch stress. 

As was mentioned in the Introduction, we base the electronic surface energy sub-model on the theories of QSE in Refs. \cite{Han,OPL}.
Thus we model $\gamma(h)$ in Pb(111) and Ag(110) films by accounting for the quantum oscillation and for the charge spilling at the film/substrate interface:
\begin{equation}
\gamma(h) = \gamma_{bf} + \frac{g_0 s^2}{(h+s)^2}\cos{\omega_1 h}\cos{\omega_2 h}-\frac{g_1 s}{h+s}.
\label{gammaQSE}
\end{equation}
The three-term form of Eq. (\ref{gammaQSE}), the oscillation, and the decay rates proportionality to $1/h^2$ and $1/h$ follow from the models based on the electron-gas (EG) 
theory  and the density functional theory (DFT); these models are described in Refs. \cite{OPL,Han}, and we refer the reader to these papers for the details. 
\emph{The form} of the oscillation in the film thickness 
is chosen to account for the beating pattern of the surface energy \cite{Kreyszig}. 
For Pb and Ag films the surface energy beating patterns were experimentally detected \cite{Smith,YJEWWNZS,Ozer,Li,CHPPM,MQA1,Gavioli} 
and computed using the EG and DFT models \cite{Han,Han1,Wu}. Eq. (\ref{gammaQSE}) with the non-oscillatory surface energy 
(i.e., $\cos{\omega_1 h}\cos{\omega_2 h}\rightarrow 1$) also has been used with some success to explain the stability and morphologies of Ag films on Si \cite{Hirayama}.
Notice that $\lim_{h\rightarrow \infty} \gamma(h) = \gamma_{bf}$ \cite{Han}, where $\gamma_{bf}=k_F^2 E_F/(80 \pi)$ \cite{OPL} is the surface energy density of a bulk film.
From Refs. \cite{Han,Han1}, $\omega_2=2\pi/\Lambda \ell$ is the frequency of the beating pattern 
and $\omega_1=2\pi/(m\lambda_F/2)$ is the primary frequency of the oscillation (within a beat).
Here $\ell$ is the interlayer spacing, or the height of 1ML of atoms,
\begin{equation}
\Lambda= \frac{\lambda_F/2}{|m\lambda_F/2-j\ell|},\qquad \mbox{(Eq. (21) in Ref. \cite{Han})}
\end{equation}
$\lambda_F=2\pi/k_F$ is Fermi wavelength, and $j>1, m$ are the smallest possible positive integers (with no common factor).
Other parameters in Eq. (\ref{gammaQSE}) are $s = 3 \pi/4k_F$ \cite{OPL} (the distance from a boundary of a quantum well in the EG model to the geometric bottom surface of a film, 
see the discussion in Refs. \cite{Han,OPL}), $g_0 = \pi E_F/36\sqrt{3}s^2$ \cite{OPL}, and $g_1$, which is the magnitude of the contribution of the charge spilling in the surface energy. 
$\gamma_{bf},\ g_0$ and $g_1$ are in units of the energy density, i.e. erg/cm$^2$.

We remark here that $g_1$ has not been measured in experiment, and there is no theoretical estimates either.
Charge spillage is necessary in quantum theories for construction of the hard-wall boundary conditions for quantum wells that contain the electron wave-function;
in turn, these conditions are indispensable for computing the surface energy oscillation \cite{Zhang,Ozer,Han1,Han,Wu}.  
We choose $g_1$ by requiring that when $\cos{\omega_1 h}\cos{\omega_2 h}=1$, all heights larger than $\ell$ (the wetting layer thickness of Pb and Ag films) are linearly unstable. 
This gives $g_1=3g_0/(1+\ell/s)\equiv g_{1c}$ \cite{JEM_Khenner}. 
Thus as far as the surface linear stability is concerned, the $g_1$ term with $g_1\le g_{1c}$ does not differentiate between various heights. 
In Sec. \ref{Ag(110)} we confirm by way of direct computation of the nonlinear evolution PDE \ref{FinalPDE} (see Appendix) that choosing $g_1\le g_{1c}$ indeed does not affect 
the emergence and stability of various preferred heights, 
and the influence of this parameter value on the island coarsening rate is very minor.

\begin{figure}[h]
\centering
\includegraphics[width=6.0in]{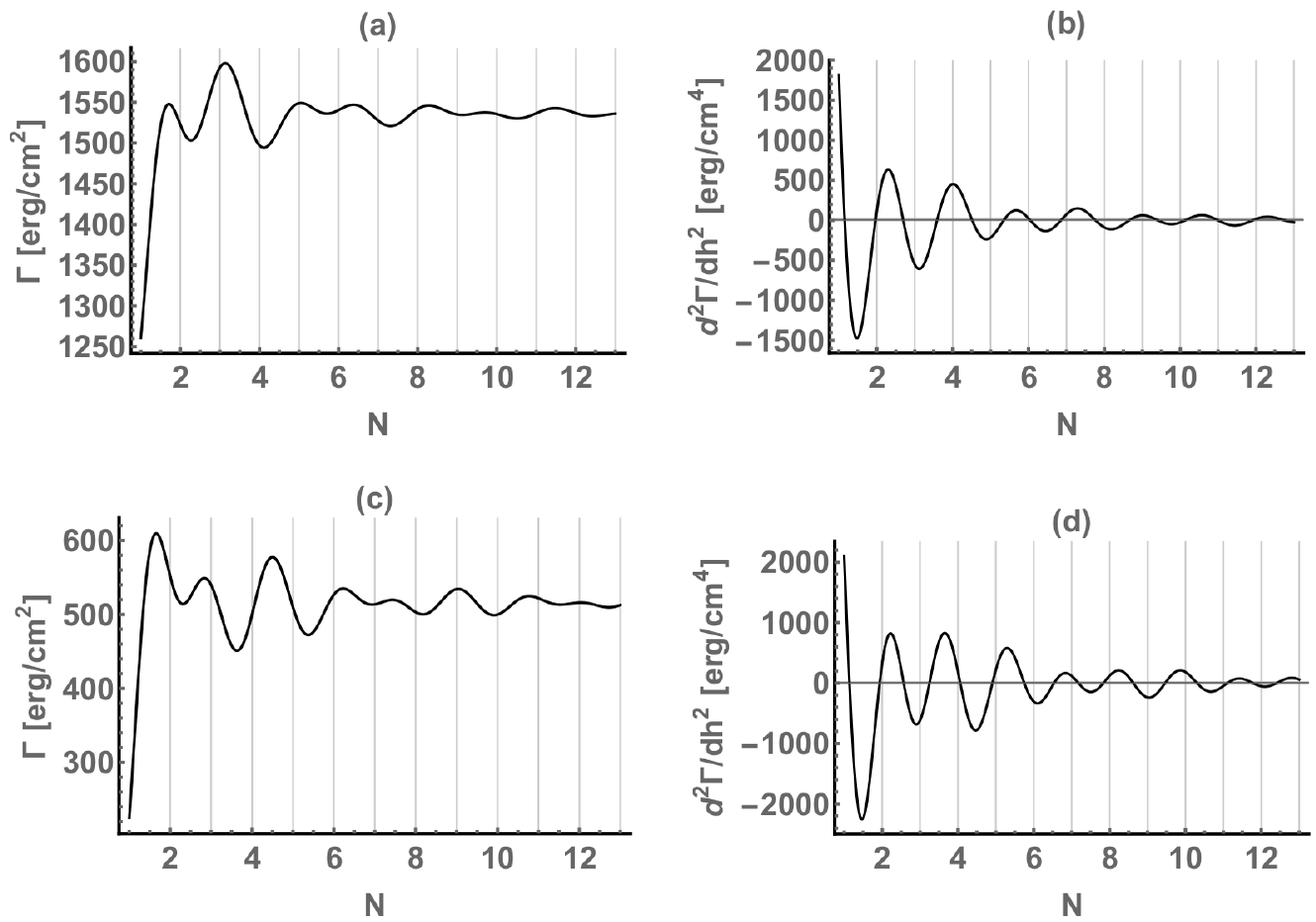}
\caption{$\Gamma(h)$ (a) and its second derivative (b) for Pb(111) surface; (c), (d): same quantities for Ag(110) surface. 
$N$ along the axis in this and other figures is the number of monolayers comprising the film height; the count includes the 1ML wetting layer.
}
\label{F1}
\end{figure}

It follows from the discussion in the last paragraph that QSE oscillation $\Gamma=\gamma_{bf} + \frac{g_0 s^2}{(h+s)^2}\cos{\omega_1 h}\cos{\omega_2 h}$ 
provides the height selection mechanism. 
$\Gamma$ and the second derivative $d^2 \Gamma/dh^2$ are shown in Fig. \ref{F1} after the film height conversion to the integer multiples
of 1ML via $h=N \ell$, where $N$ is the number of monolayers. These plots can be compared to the similar 
plots in Refs. \cite{Han,Han1,Ozer,Wu}.
In order to underscore the continuum nature of the model, all integer monolayer heights are connected by the solid lines in Fig. \ref{F1} 
and all other Figures in the paper.
From the plots of the second derivative the stable 
film heights can be read off, since a stable height corresponds to a positive value of $d^2\Gamma/dh^2$ \cite{Zhang,Ozer,Han1,Han,Wu}. In Table \ref{Tab2} we 
provide, for reference, the stable heights sets obtained by various methods. These sets display a bilayer oscillation, which is interrupted at a certain monolayer height.
The oscillation then resumes from the next monolayer height, and this feature cyclically repeats until the oscillation dies off
in the relatively thick films ($h\sim 25\div 30$ML from the cited theories and the experiments \cite{Ozer}). In the dynamical setting of the evolving nanoisland morphologies,
a stable island height may be truly stable, or it may be metastable.

Values of $E_F,\ k_F,\ \ell,\ j,\ m$, as well as the film's shear modulus $\mu_F$, its Poisson's ratio $\nu_F$, and the (dimensionless) lattice misfit $\epsilon$ 
are shown in Table \ref{Tab1}. 
These values and the above-stated definitions of $g_0$, $g_1$, $\omega_1$, $\omega_2$, $\gamma_{bf}$, $s$ establish values of
seven dimensionless parameters $G_0$, $G_1$, $\Omega_1$, $\Omega_2$, $\mathcal{E}$, $E_{0013}$, and $E_{0004}$ that enter the dimensionless evolution equation. 
The first four parameters in the latter set are surface energy-related, and the last three parameters (all defined in Appendix) are stress-related.
The only variable in the simulations is the random initial island distribution.

The last two terms in Eq. (\ref{gammaQSE}) are responsible for wetting effect. The term $-g_1 s/(h+s)$ provides the long-range surface-substrate attraction, 
while the term 
$(g_0 s^2/(h+s)^2)|\cos{\omega_1 h}\cos{\omega_2 h}|$ provides the shorter-range repulsion. 
The repulsion would result in a wetting layer of a thickness of the order 
of a "wetting length" $s$ \cite{Chiu,Golubovic}. A wetting layer prevents the film from dewetting the substrate. Using values from Table \ref{Tab1}, from the above stated 
expression for $s$ we obtain $s=1.48 \angstrom$ (Pb(111)) and $s=1.95 \angstrom$ (Ag(110)).
These values are indeed of the order of 1ML, the correct wetting layer thickness. van der Waals attraction and repulsion is typically modeled by the same two terms in $\gamma(h)$
\cite{Golubovic}, only the van der Waals energies (Hamaker constants) replace $g_0$ and $g_1$ and $\cos{\omega_1 h}\cos{\omega_2 h}\rightarrow 1$. Since for ultrathin 
films van der Waals energies are several orders of magnitude 
smaller than their electronic counterpart $E_F$ \cite{OPL}, the van der Waals contributions to wetting are not considered in this paper.

%
\begin{table}[!ht]
\centering
{\scriptsize 
\begin{tabular}
{|c|c|c|c|}

\hline
				 
			\rule[-2mm]{0mm}{6mm} \textbf{Parameter} & \textbf{Description}	 & \textbf{Pb(111)}  & \textbf{Ag(110)}\\
			\hline
            \hline
			\rule[-2mm]{0mm}{6mm} $E_F$ & Fermi energy & 9.584 eV ($1.535\times 10^{-11}$ erg) & 5.55 eV ($8.89\times 10^{-12}$ erg)   \\
            \hline
			\rule[-2mm]{0mm}{6mm} $k_F$ & Fermi wavenumber & $1.586 \angstrom^{-1}$ ($1.586\times 10^8$ cm$^{-1}$) & $1.207 \angstrom^{-1}$ ($1.207\times 10^8$ cm$^{-1}$)   \\
			\hline
     		\rule[-2mm]{0mm}{6mm} $\ell$ & Interlayer spacing ($=$1ML height) & $2.84 \angstrom$ ($2.84\times 10^{-8}$ cm) & $1.44 \angstrom$ ($1.44\times 10^{-8}$ cm)    \\
			\hline
			\rule[-2mm]{0mm}{6mm} $j$ & & 2 & 2     \\
			\hline
			\rule[-2mm]{0mm}{6mm} $m$ & & 3 & 1     \\
			\hline
			\rule[-2mm]{0mm}{6mm} $\mu_F$ & Shear modulus & $56\times 10^9$ erg/cm$^3$ & $300\times 10^9$ erg/cm$^3$     \\
			\hline
			\rule[-2mm]{0mm}{6mm} $\nu_F$ & Poisson's ratio & 0.44 & 0.37     \\
			\hline
			\rule[-2mm]{0mm}{6mm} $\epsilon$ & Lattice misfit & 0.05 & 0.01     \\
\hline
				
\end{tabular}}
\caption[\quad Physical parameters]{Physical parameters (from Ref. \cite{Han}). 
Using values from the Table, $\Lambda=7.55$ for Pb(111) and $\Lambda=9.39$ for Ag(110) \cite{Han1,Han}.}
\label{Tab1}
\end{table}
\begin{table}[!ht]
\centering
{\scriptsize 
\begin{tabular}
{|c|c|c|}

\hline
				 
			\rule[-2mm]{0mm}{6mm} \textbf{Method} & \textbf{Pb(111)}  & \textbf{Ag(110)}\\
			\hline
            \hline
			\rule[-2mm]{0mm}{6mm} EG, for free-standing film & 2, 4, 5, 7, 9, 11, 12, 14, 16, 18, 19, 21, 23 \cite{Han} & 2, 4, 5, 7, 9, 11, 13, 16, 18 \cite{Han}   \\
            \hline
			\rule[-2mm]{0mm}{6mm} DFT, for free-standing film & 2, 4, 6, 8, 9, 11, 13, 15, 18, 20, 22, 24 \cite{Han} & 2, 5, 7, 9, 11, 13, 16, 18 \cite{Han}   \\
			\hline
			\rule[-2mm]{0mm}{6mm} This work, for free-standing film  & 2, 4, 6, 7, 9, 11, 12, 14, 16, 17, 19  & 2, 4, 5, 7, 8, 10, 13, 16, 19    \\
            \hline
            \rule[-2mm]{0mm}{6mm} Experiment & 3, 6, 8, 10(11), 15, 17(18), 20(22) (on Cu(111) \cite{OVM,DKGTH});  & 2, 4, 6, 8?, 10?,... (on NiAl(110) \cite{Han1})    \\
			\rule[-2mm]{0mm}{6mm}            & 6, 8, 10, 12, 14(15), 17, 19, 21 (on Si(111), cited in Ref. \cite{Ozer}, p. 229) & 
			\\

\hline
				
\end{tabular}}
\caption[\quad Stable heights]{Stable island heights (in monolayers). The monolayer count includes the 1ML wetting layer.}
\label{Tab2}
\end{table}

It should be remarked here, that in the case of homoepitaxial single-crystal film growth (e.g., Ag on Ag(110)) the anisotropy of adatoms diffusivity leads to formation
of anisotropic islands (square or hexagonal mounds) and ripples of different orientation \cite{MFHL,RBFHV,MCBV,GAMBFV,VHF}. Anisotropy of the diffusivity arises due to the 
intrinsic anisotropy 
of the substrate. However, 2D heteroepitaxial islands that result from ``quantum" growth method do not seem to exhibit systematic anisotropic shapes 
\cite{Ozer,Han1,CHPPM,SCT,WMJFJJX}; this suggests
that the anisotropies of the diffusivity and of the surface energy do not significantly affect the morphologies. Indeed, we did not find any mention of either anisotropy 
in the just cited works and 
in other major experimental or modeling works on quantum growth that are cited in the Bibliography. Recent modeling \cite{DDM} of the morphological evolution of surfaces of stressed 
solids suggests that, at least in the framework of 2D model, the anisotropy of the diffusivity does not significantly affect the dynamic morphologies (unless
surface electromigration is present). Also, computations using simpler version of our model \cite{MSMSE_Khenner} show that the surface energy anisotropy 
is not making a large impact on the 2D (metal) islands morphologies (for growth and coarsening of 3D metal islands the insignificance of the surface energy anisotropy 
was noted, see Ref. \cite{L}.)

To compute the dynamics of the quantum nanoislands in the next section, we use the dimensionless form of a ``thin-film expansion" of Eq. (\ref{MullinsEq}). 
The details of the reduction of Eq. (\ref{MullinsEq}) to its thin-film form, adimensionalization, and values of the physical parameters involved in calibrating 
the equation and the computational domain can be found in Appendix.

\section{Computations of quantum nanoislands dynamics}
\label{EvolveIslands}

In this section we numerically study the emergence of nanoislands with the ``magic" heights and their coarsening dynamics and metastability. 
The final computed morphologies contain either one island (the equilibrium) or two islands (near-equilibrium). (Since the dynamics slows with the passage of time, to save
the computing time, quite often we chose to abort the computation prior to reaching the equilibrium configuration of a single island.)

\begin{figure}[h]
\centering
\includegraphics[width=6.0in]{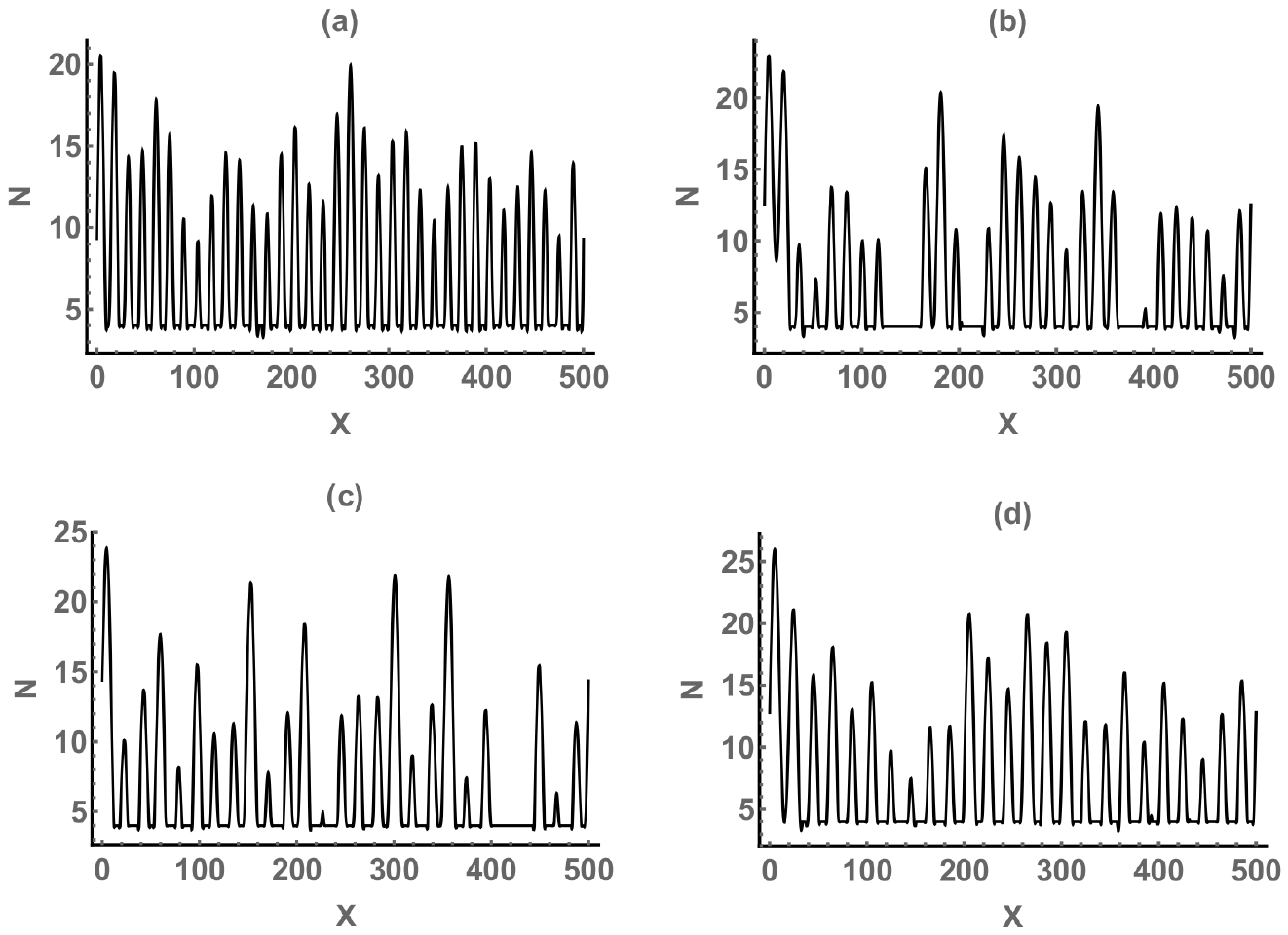}
\caption{Examples of the initial conditions used for computations of the islands dynamics.
}
\label{F2}
\end{figure}

Fig. \ref{F2} shows the examples of the initial conditions used in the simulations. As can be seen from Fig. \ref{F2}, we independently and randomly vary the heights, spacings, 
and the number (and thus the width) of the non-electronic (non-quantum) islands comprising the initial condition. (Due to a vast difference of the horizontal and vertical scales, 
see Appendix, the rounded islands' tops appear as spikes.) 
The islands ``grow out" of the base film that is 4ML thick. 
Although Su et al. \cite{SCT} do not find a flat-top Pb islands with the thickness below 4ML, our choice of the 4ML base thickness is due to the computational reasons only, 
as we found by experimentation that this base thickness is the smallest value for which the computation is stable and verifiable by a computational grid refinement. 
The number of islands varies between 20 and 40, and their height is restricted by 20ML (above the 4ML base film thickness).
Notice that, again because of the continuum nature of the model, all non-integer and integer monolayer heights will be computed using Eq. (\ref{FinalPDE}) 
shown in Appendix.
%

\subsection{Ag(110)}
\label{Ag(110)}

Fig. \ref{F3} shows several snapshots of the computed surface taken from \emph{Ag(110)Movie-Trial1} (see Supplementary Material). Such sequence of morphology transitions is typical 
for all simulations of Ag(110) nanoislands, though the details may be somewhat different. Prior to discussing this figure, we want to point out that the features of the islands 
that appear curved, or rounded, remain curved or rounded at larger magnifications and no matter how large or small are the islands, as can be seen from Fig. \ref{F8}(f) that we generated specifically
to demonstrate this.

In Fig. \ref{F3}, in the beginning, the initial islands set coarsens very fast into a smaller
set of about 10 islands (Fig. \ref{F3}(b)); some of these islands already have the ``electronic" height 8, 10, or 13ML and the flat top (see Table \ref{Tab1}), 
other islands have the non-electronic height and/or are rounded at the top. Also at this stage the islands choose their base, i.e. the apparent film thickness they grow out from; 
this thickness always has the electronic value 4 or 6ML. This matches the experimental observation \cite{Smith,YJEWWNZS} that the flat film morphology can develop only in a 
fully wetting layer for a thickness of more than 6ML. With the increasing time, the islands undergo further coarsening by, first, eliminating all non-electronic heights (Fig. \ref{F3}(d)). 
After this is completed, the process of steady elimination of the metastable (i.e., the smallest) electronic heights is started. In the case of Fig. \ref{F3}(d) the metastable electronic 
heights are 8 and 10ML, and the smallest 8ML height is eliminated first. 8ML islands shrink laterally by transferring mass to the adjacent 10ML island(s), see the evolution of the 
third island from the left in Figures \ref{F3}(c,d,e); in this process the
flat electronic top is preserved until the island area becomes sufficiently small. Then the top becomes rounded and from that point on the island height decreases simultaneously with 
its area until the island disappears (Figures \ref{F3}(e,f)).
8ML height also can be ``squeezed out"; this happens if this height occurs in the form of a terrace separating two 10ML islands, see the central section of the film in Figures \ref{F3}(d,e,f).
After only 10 and 13ML electronic heights remain (Fig. \ref{F3}(g)), the elimination of 10ML islands begins similar to the just described elimination of the 8ML islands.
Notice how in Figures \ref{F3}(g,h) the second and the third 10ML islands squeeze out the 6ML terrace between them and become a single 10ML island, while the smallest (third) 
10ML island is absorbed into the last 13ML island (the fourth, also the first due to periodicity of the computational domain). In Fig. \ref{F3}(i) only one 10ML and one 13ML islands remain; 
the 10ML island will be ultimately absorbed into the 13ML island and the evolution will stop. Overall, Fig. \ref{F3} clearly shows that on the path to the equilibrium, 
the evolution of Ag nanoislands is in the 2D growth mode for nearly all of the time.
\begin{figure}[h]
\centering
\includegraphics[width=6.0in]{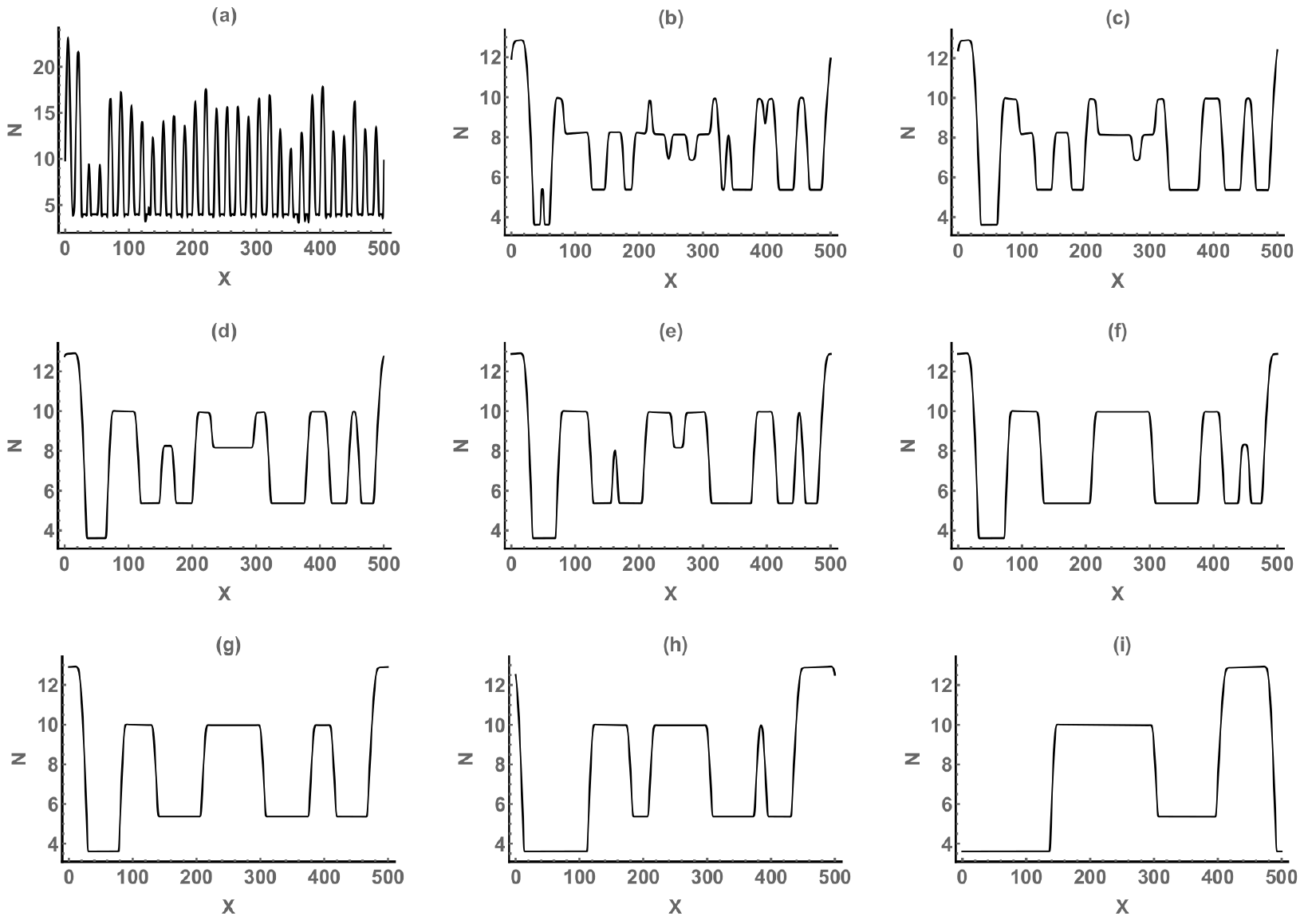}
\caption{Snapshots from \emph{Ag(110)Movie-Trial1} at (a) $T=0$, (b) $T=2\times 10^3$, (c) $T=6\times 10^3$, (d) $T=2\times 10^4$, (e) $T=5\times 10^4$, (f) $T=7.5\times 10^4$, (g) $T=1.25\times 10^5$, 
(h) $T=5\times 10^5$, and (i) $T=10^6$.
}
\label{F3}
\end{figure}
%

\begin{figure}[!ht]
\centering
\includegraphics[width=4.0in]{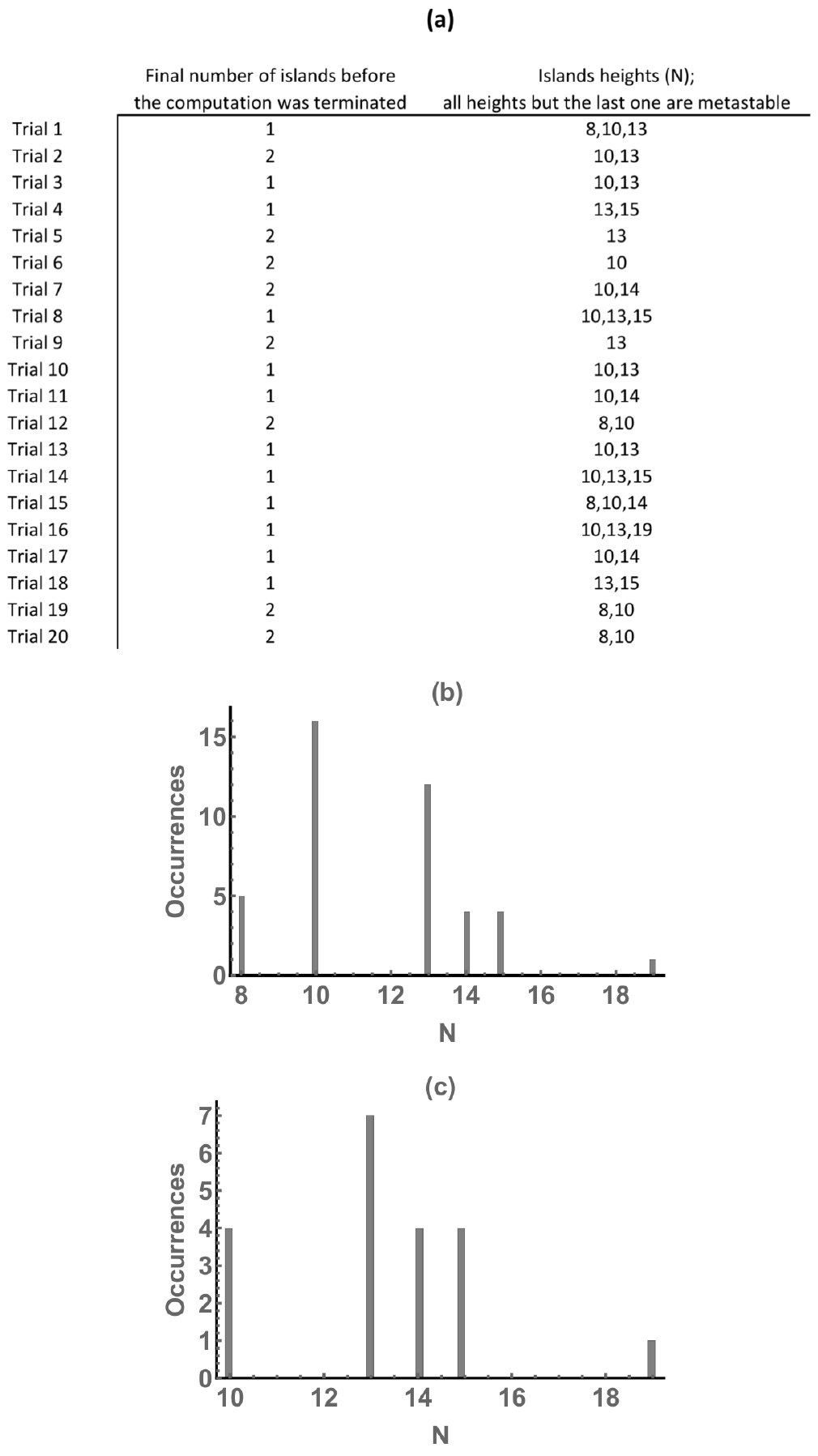}
\caption{(a) Raw statistical data on the flat-top islands, collected from the numerical experiments with Ag(110) film; (b) Histogram plot of all heights in (a); 
(c) Histogram plot of the stable heights in (a). Data in (a) labeled as ``Trial 1", ``Trial 3",
``Trial 9" were collected from \emph{Ag(110)Movie-Trial1}, \emph{Ag(110)Movie-Trial3}, and \emph{Ag(110)Movie-Trial9}.}
\label{Tab3}
\end{figure}

In Fig. \ref{Tab3} we summarize the data from our computational experiments with Ag(110) films. It is clear from the presented Table that the electronic heights 8, 10, and 13ML are
prevailing. Non-electronic 14 and 15ML heights occurred a few times only, and the electronic 19ML occurred once. It is also seen that 10 and 13ML are favored over 8ML, and 6ML 
never occurred, as this electronic height always serves as the base for taller islands, as in Fig. \ref{F3}. Another observation is that 8ML islands seem to always
occur alongside with 10ML and 13ML islands, while these islands infrequently do occur as the only island species throughout the simulation, as in the Trials 5, 6 and 9.
The reason for the smaller than expected fraction of the metastable 8ML islands is most likely the computational bias introduced by the simulation's initial condition, 
rather than the physical. Indeed, the initial conditions in Fig. \ref{F3} (that we chose among others like them to show the visually quite different cases) display fewer 
8ML initial heights than another heights, thus there is a lower probability of the 8ML island to occur later in the evolution.

\begin{figure}[h]
\centering
\includegraphics[width=6.0in]{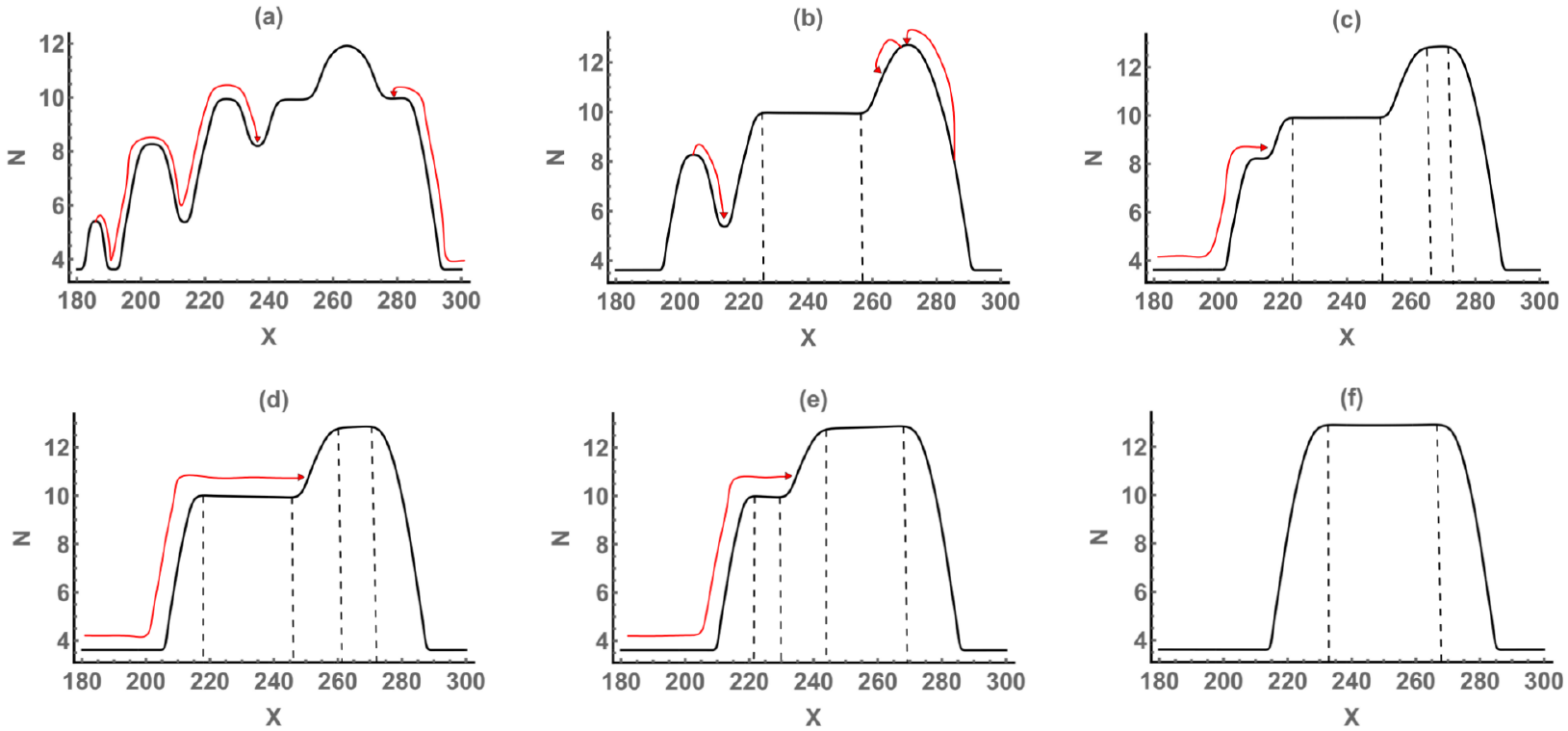}
\caption{(Color online.) Zoomed snapshots from \emph{Ag(110)Movie-Trial9} at (a) $T=2\times 10^3$, (b) $T=6\times 10^3$, (c) $T=1.2\times 10^4$, (d) $T=1.6\times 10^4$, (e) $T=2.8\times 10^4$, 
and (f) $T=3.4\times 10^4$.
}
\label{F4}
\end{figure}

\begin{figure}[h]
\centering
\includegraphics[width=6.0in]{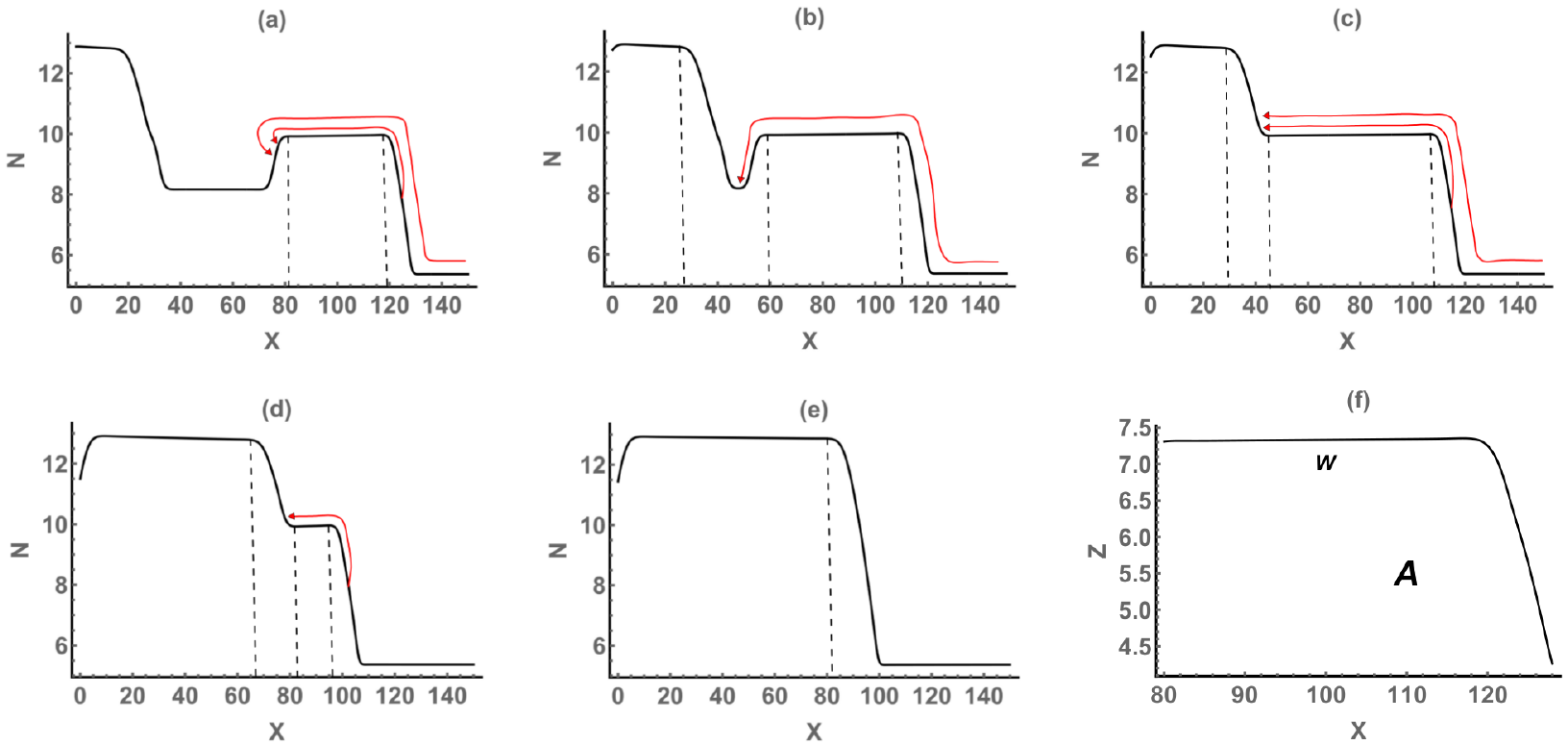}
\caption{(Color online.) Zoomed snapshots from \emph{Ag(110)Movie-Trial3} at (a) $T=4\times 10^4$, (b) $T=8.5\times 10^4$, (c) $T=10^5$, (d) $T=1.65\times 10^5$, and (e) $T=2\times 10^5$.
(f) is the right island enlarged from (a), where the vertical axis is converted from 
monolayers to the conventional dimensionless height, 
as described in Appendix. $W$ marks the width of the top terrace, and $A$ the island area.
}
\label{F6}
\end{figure}

Next, in Figures \ref{F4} and \ref{F6}, we catalog the major morphological transitions that are experienced by a single Ag island on its way to a final
stable electronic height. These snapshots provide a unique understanding of the sequence of the mass transfers by the ``electronically-biased" surface diffusion that culminate
in the stable electronic height. To our knowledge, such level of detailing was not so far achieved neither by the in-situ microscopy in the experiment \cite{YJEWWNZS,UFTE,Han1,Li,Hirayama,CHPPM,MQA1,Gavioli},
nor by the models \cite{Ozer,UFTE,Li,Kuntova}. Red arrows in these figures mark the origin, direction, and destination of the atoms surface diffusion transport.

In the right half of Fig. \ref{F4}(a) a non-electronic 12ML island is starting to transform into the electronic 13ML island. 12ML island has the rounded cap bounded by the 
narrow horizontal terraces at the electronic 10ML height. The left terrace is adjacent to three smaller islands. The atoms comprising the smallest (first) of these islands
are transported to the trough adjacent to the left terrace, filling it completely and elongating this terrace, so that the closest to the terrace small island is 
absorbed into the large island, see Fig. \ref{F4}(b). Simultaneously, the atoms from the smaller islands to the right of the large island (not shown) diffuse along the flat 
section of the film at 4ML and then upward along the island wall,
onto the right terrace and then to the cap; the cap grows to the right and vertically by 1ML (reaching the electronic height 13ML), until the terrace disappears and the 
smooth, slightly curved right wall emerges, as seen in Fig. \ref{F4}(b).

In Fig. \ref{F4}(b), the material in the remaining small island is used to fill the trough; Fig. \ref{F4}(c) shows the moment when the small terrace is created below the 
large terrace. (If a 3D extension of Fig. \ref{F4}(c) is looked from above, this structure resembles the familiar ``wedding cake".) Also, some atoms from the curved right wall
are transported to the island rounded top and then to the left wall of the cap; some atoms from the rounded cap also diffuse to the left wall of the cap. This creates the flat island
top at 13ML and the straight right island wall, as seen in Fig. \ref{F4}(c).

In Figures \ref{F4}(c-e) the upward transport of atoms (from the smaller islands on the left of the featured island, not shown) along the straight left island wall creates a wide flat top terrace at 13ML, while the terrace at 10ML steadily shrinks;
this process finally results in the final, roughly trapezoidal island shape shown in Fig. \ref{F4}(f).

Fig. \ref{F6} shows the consumption of smaller electronic 10ML island by larger electronic 13ML island. Again the essential diffusion transport processes are the upward transport 
of the material from the adjacent small islands along the island wall; the wall erosion and retraction by removing some material and transporting it upward; 
and diffusion along the terrace
and deposition of atoms in the trough or on the opposite wall, which extends the island laterally without alteration of the shape.

\begin{figure}[h]
\centering
\includegraphics[width=6.0in]{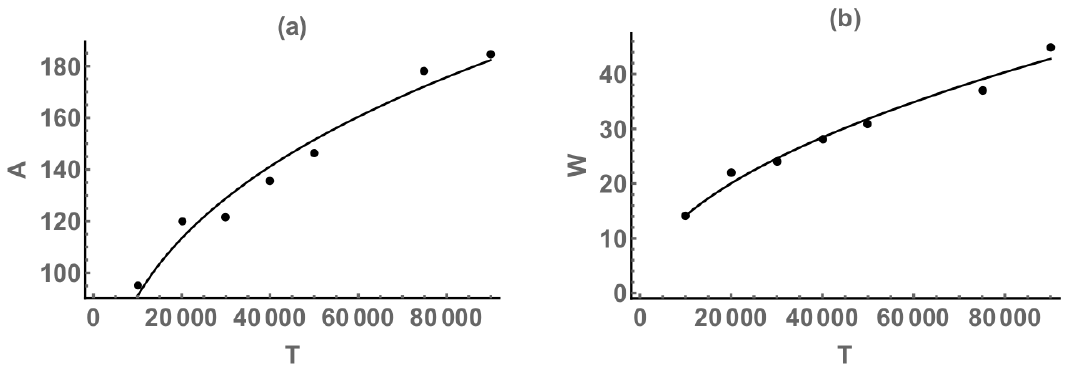}
\caption{The island area (a) and the width of the island's flat top terrace (b) as functions of the dimensionless time, for a typical metastable 
growing island on Ag(110) surface. Power-law fits are shown by the solid curves:
(a) $A=4.98\ T^{0.316}$,  (b) $W=0.135\ T^{0.505}$.
}
\label{F5_1}
\end{figure}

Fig. \ref{F5_1} shows the area of the right island in Fig.  \ref{F6}(a-c) and the width of its flat top terrace as functions of time.
We considered the time interval during which the island enlarges and before it merges with the left island. 
The area scales as $A=4.98\ T^{0.316}\sim T^{1/3}$ and the width as $W=0.135\ T^{0.505}\sim T^{1/2}$. Both scaling exponents are common in various growth and coarsening 
phenomena \cite{RV,C,Alik}, but for electronic growth they are noted here for the first time.

\begin{figure}[h]
\centering
\includegraphics[width=3.0in]{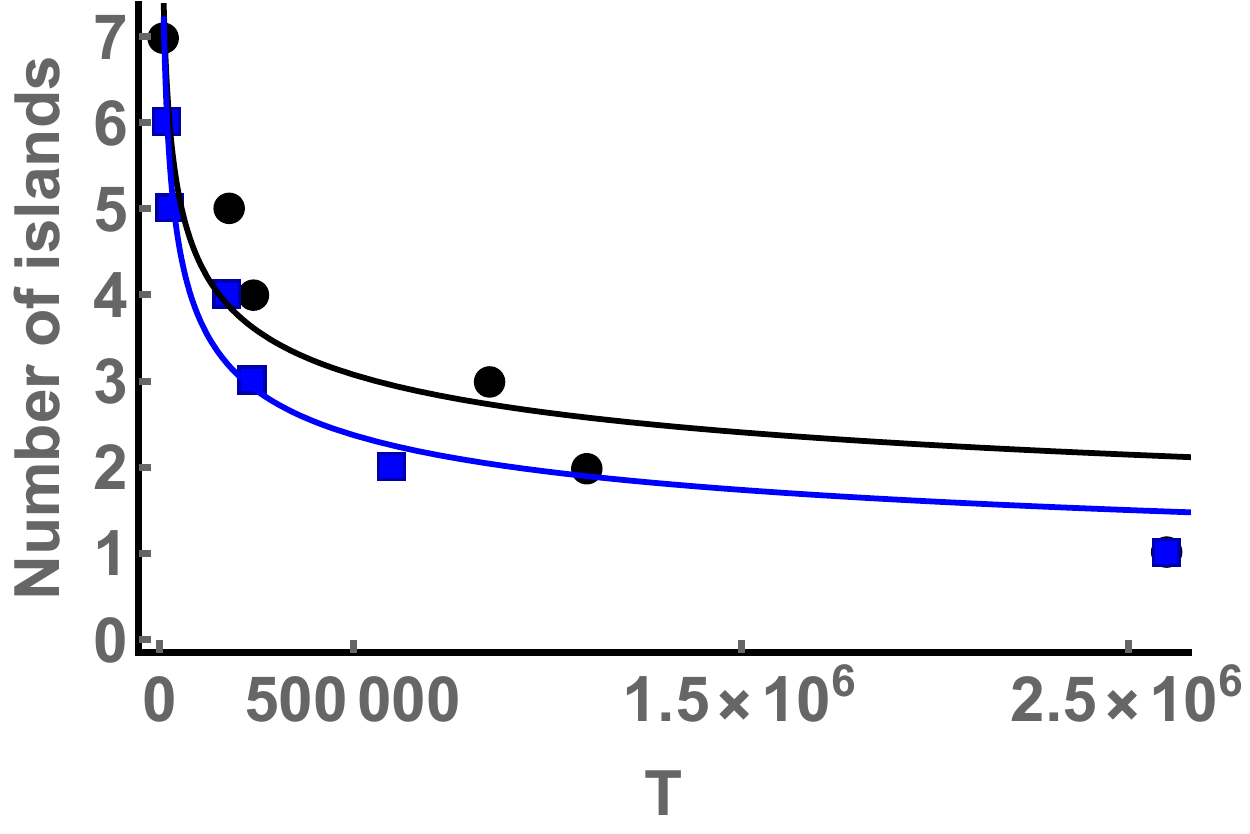}
\caption{(Color online.) The number of flat-top and round-top islands, NI, on Ag(110) surface vs. the time. Squares: Trial 3, Circles: Trial 16. 
Power-law fits are shown by the solid curves: NI$=98.16\ T^{-0.284}$ (Trial 3),  NI$=57.4\ T^{-0.223}$ (Trial 16). Data for the plot is partially taken from
\emph{Ag(110)Movie-Trial3}.
}
\label{F7}
\end{figure}

Fig. \ref{F7} shows the number of electronic (flat-top) and non-electronic (rounded top) Ag islands vs. the time, for two representative trials.
The fits are NI$=98.16\ T^{-0.284}$ (Trial 3) and 
NI$=57.4\ T^{-0.223}$ (Trial 16), where NI stands for ``Number of Islands". Thus the typical length scale $\langle L\rangle$ (the distance between the islands) is roughly 
estimated by the averaged reciprocal value, i.e. $\langle L\rangle \sim T^{1/4}$. The exponent 1/4 is typical for coarsening driven by the surface diffusion (in the absence of the
lattice-mismatch stress) \cite{C,Alik,SGDNV,OL}.

\begin{figure}[h]
\centering
\includegraphics[width=6.0in]{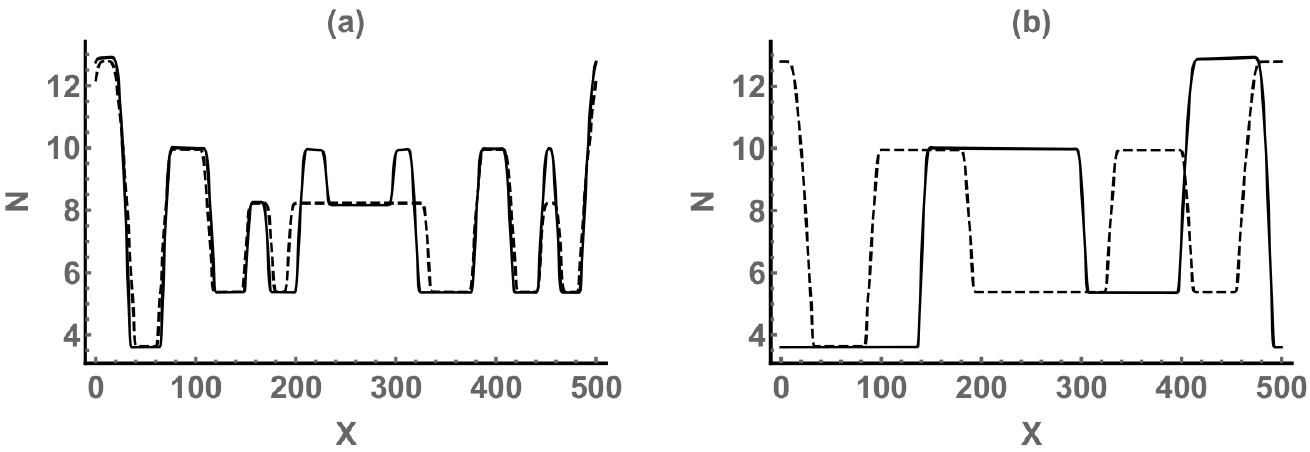}
\caption{Ag(110) surface at $g_1=g_{1c}=3g_0/(1+\ell/s)$ (solid lines) and $g_1=g_{1c}/15$ (dashed lines). 
(a) $T=2\times 10^4$, (b) $T=10^6$. The initial condition in Fig. \ref{F3}(a) was used. Notice that $g_{1c}$ is used to compute other Figures in the paper.
}
\label{F7_1}
\end{figure}

Finally, in Fig. \ref{F7_1} we compare the surface morphologies at various values of the amplitude $g_1$ of the charge spilling contribution in the surface energy 
(see Eq. \ref{gammaQSE}). 
These computations were executed with the same initial condition.
By decreasing $g_1$ fifteen times the coarsening is made slightly faster at short times, but later it slows, and in the end of the simulation just one extra island remains for the 
small $g_1$ case. Thus the effect of $g_1$ variation on coarsening is not strong, and we did not attempt to quantify it. Also, in reference to the discussion in
Sec. \ref{Formulation} about the impacts of $g_1$ on the height selection, we computed the evolution with this  
initial condition and other $g_1$ values smaller than $g_{1c}$. These computations confirmed that the emergence and metastability
of preferred heights is not affected by these values; that is, for Trial 1 ``magic" heights are 8, 10, and 13 ML at $g_1\le g_{1c}$, see Fig. \ref{Tab3}.
Similar conclusions are expected to fully hold for other computations in the paper.

In the summary of this section, we remark that the typically small values of the lattice mismatch for Ag(110) films grown on metals such as Fe ($\epsilon=1\%$ \cite{JP}, 
the value used in our computations) allow the electronic effects to dominate in the evolution of the film morphology. This leads to the prevalence of the islands with the distinct 
electronic shape, i.e. a trapezoid, which is reinforced by the special sequence of the surface diffusion events as we discussed above. 
(In 3D, the expected island shape is a truncated cone or a truncated pyramid.) The upward diffusion on the island sidewall, being, as we discussed, the important mechanism of the 
morphology evolution, was noted previously in the Monte Carlo computations of Pb nanoislands on silicon \cite{Kuntova} and recently, in  the Kinetic Monte Carlo computations
of the metal islands on graphene \cite{LAGGS}. 
Results of our modeling are clearly in-step with STM studies of quantum nanoislands in Refs. \cite{UFTE,CHPPM,Gavioli,SCT}. In particular, see Fig. \ref{F3}(g-i), once the 
selected height of a large-area island  ``is achieved, then this height is strongly favored in subsequent growth (i.e., lateral spreading of a mesalike island bounded by 
relatively steep side walls)" \cite{UFTE}.

%
%
%
%
%
%

\subsection{Pb(111)}
\label{Pb(111)}

In this section, we turn attention to modeling the quantum morphological evolution of an ultrathin Pb(111) films. The lattice mismatch of Pb films grown on copper or silicon is very large (about 10\% \cite{HHTWHT}). 
With our stress model, we are able to compute the morphological evolution at the maximum $5\%$ lattice-mismatch strain $(\epsilon=0.05)$. 
All results that we show in this section correspond to this $\epsilon$ value. (At $\epsilon > 0.05$ the strongly nonlinear terms proportional to $\mathcal{E}$ in the 
evolution PDE (\ref{FinalPDE}) hinder the computation over the long time span of the morphological evolution to the equilibrium. The integration of these stiff terms takes too 
long with our available computational resources.)

\begin{figure}[!ht]
\centering
\includegraphics[width=4.0in]{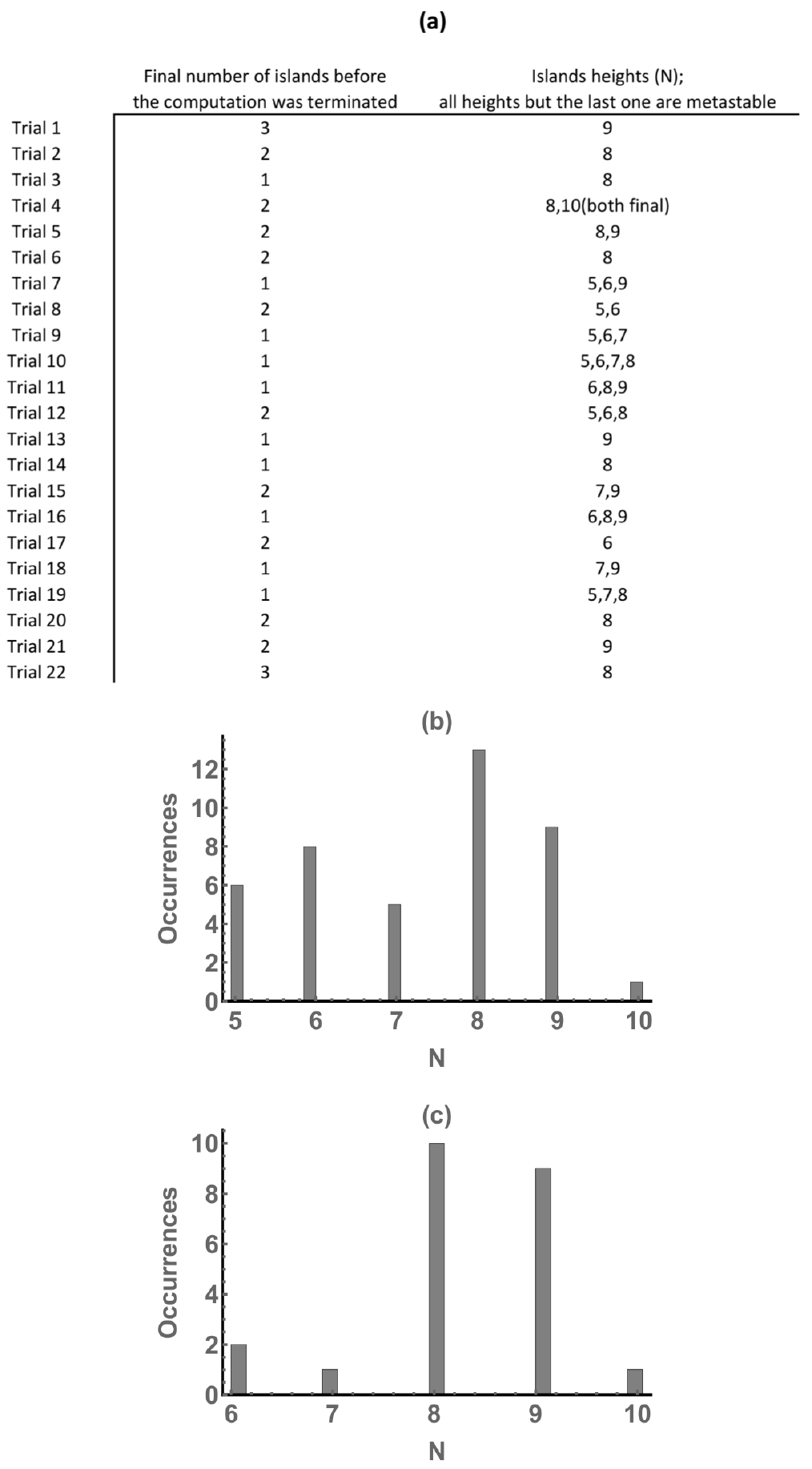}
\caption{(a) Raw statistical data on the flat-top islands, collected from the numerical experiments with Pb(111) film; (b) Histogram plot of all heights in (a); 
(c) Histogram plot of the stable heights in (a). Data in (a) labeled as ``Trial 16" were collected from \emph{Pb(111)Movie-Trial16}.}
\label{Tab4}
\end{figure}

The data in Fig. \ref{Tab4} shows that 6, 8, and 9ML islands appear most often. 6ML and 8ML islands typically emerge in the experiment, see 
Table \ref{Tab2}, while 9ML islands are the feature of our stress-free electronic sub-model and of the EG model (also stress-free). Thus the set of the most probable islands heights
in Fig. \ref{Tab4} is the combination that reflects the stress influence on the quantum heights. Also, in most trials, see the trials 4, 11, 12, 15, 16, 18, 19, the characteristic 
bilayer oscillation imposed by the quantum size effect can still be observed. Via a variable temperature STM, Calleja et al. \cite{CHPPM} observed that 8ML islands are most stable at various temperatures. Our computational 
results shown in Fig. \ref{Tab4}(b,c) correlate well with this observation.

\begin{figure}[h]
\centering
\includegraphics[width=6.0in]{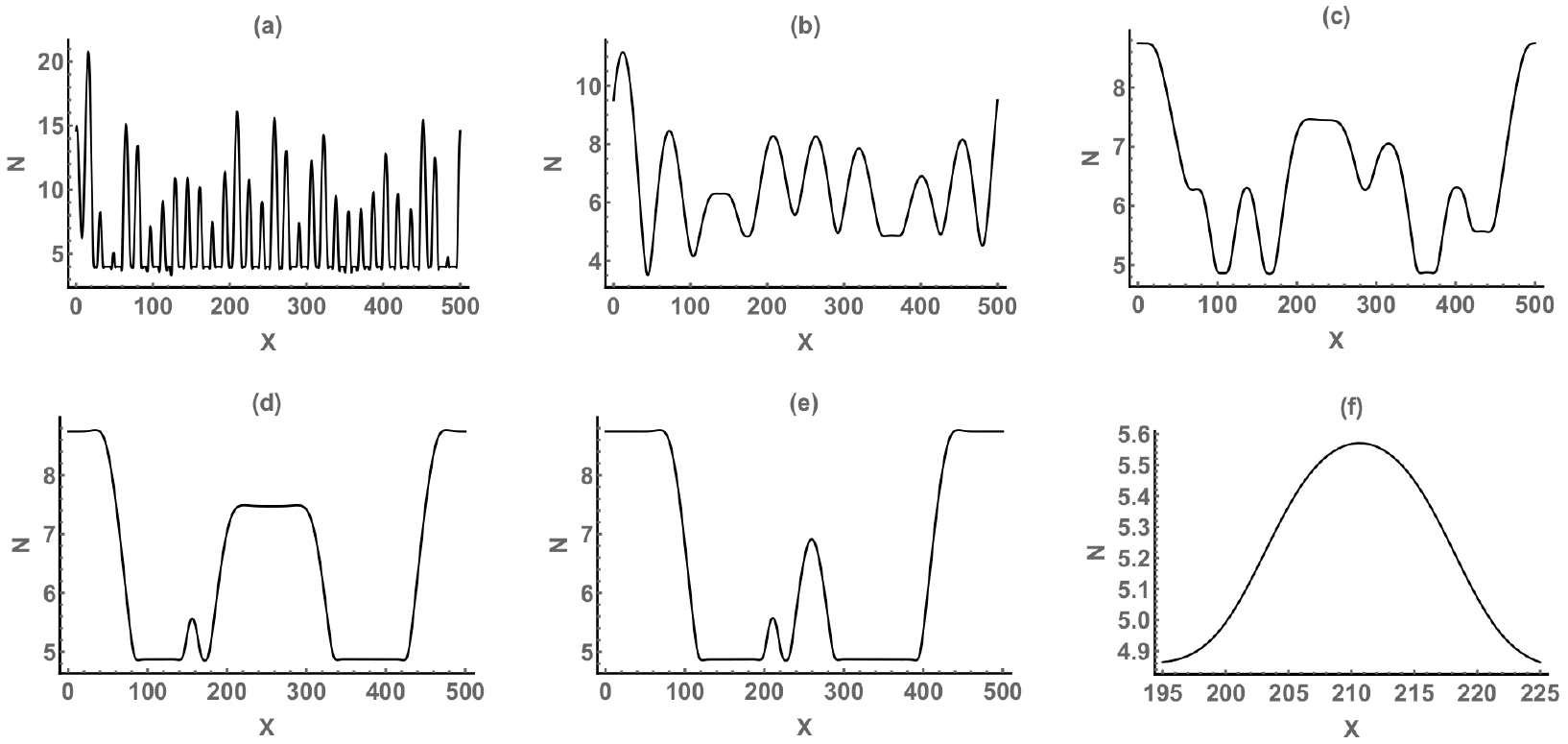}
\caption{Snapshots from \emph{Pb(111)Movie-Trial16} at (a) $T=0$, (b) $T=3\times 10^3$, (c) $T=2\times 10^5$, (d) $T=1.2\times 10^6$, (e) $T=8.5\times 10^6$. (f) is the zoom into the smallest island in (e), showing 
its typical Stranski-Krastanow shape \cite{SFAA}.
}
\label{F8}
\end{figure}

In Fig. \ref{F8} the typical morphological evolution is shown. One can notice, at various stages, the formation of 6ML and 8ML metastable islands; only the 9ML islands are 
present in the final computed morphology. Overall, there is more rounded islands at any stage than the flat-top islands, which is the opposite of Ag(110), see Fig. \ref{F3}. 
As we remarked in the Introduction, the rounded islands are typical for Stranski-Krastanow growth \cite{SFAA}, where the lattice-mismatch stress is the dominant effect in 
the morphology evolution. This explains the prevalence of such islands on the Pb(111) surface.

\begin{figure}[h]
\centering
\includegraphics[width=6.0in]{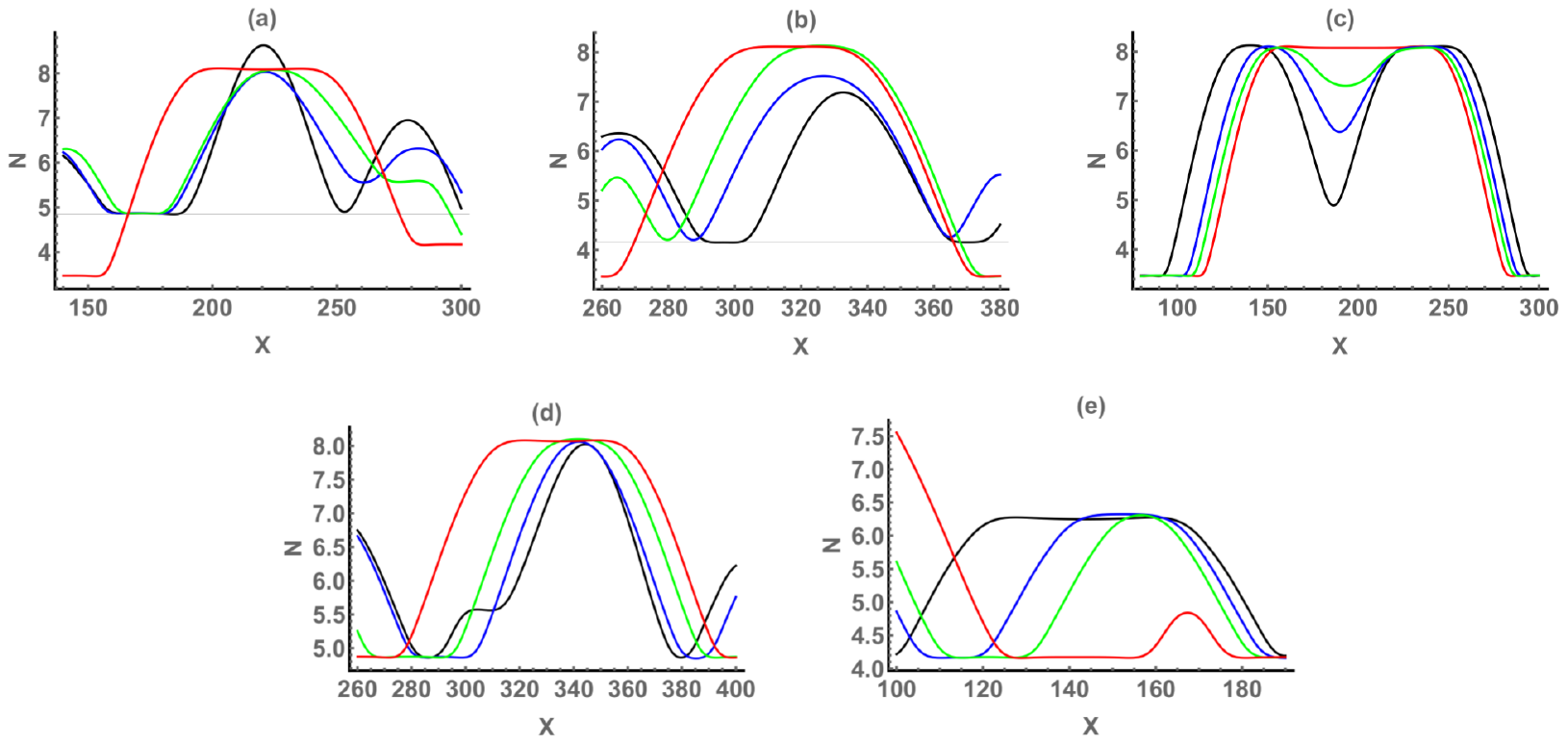}
\caption{ (Color online.) Zoom into a section of Pb(111) surface from different simulations, showing (a-d): various scenarios of a formation of the stable, flat-top 8ML island, and (e): the shrinking and disappearance of a metastable 6ML island. 
Black, blue, green, and red colors mark the increasing time.
}
\label{F9}
\end{figure}

Fig. \ref{F9}(a-d) shows how a stable flat-top Pb island may be formed. We notice immediately that these scenarios are very different in the presence of the large mismatch stress,
compare to Figures \ref{F4} and \ref{F6}. At a glance, a stable electronic island is formed by smoothly transforming a non-electronic (i.e., rounded) island. Such island
grows laterally and simultaneously grows or shrinks vertically, until its top aligns with the electronic height. Next, the top flattens out and the resulting trapezoid-shaped 
island grows more laterally.  Disappearance of the metastable electronic islands is through the inverted sequence of steps, see Fig. \ref{F9}(e).

\begin{figure}[h]
\centering
\includegraphics[width=6.0in]{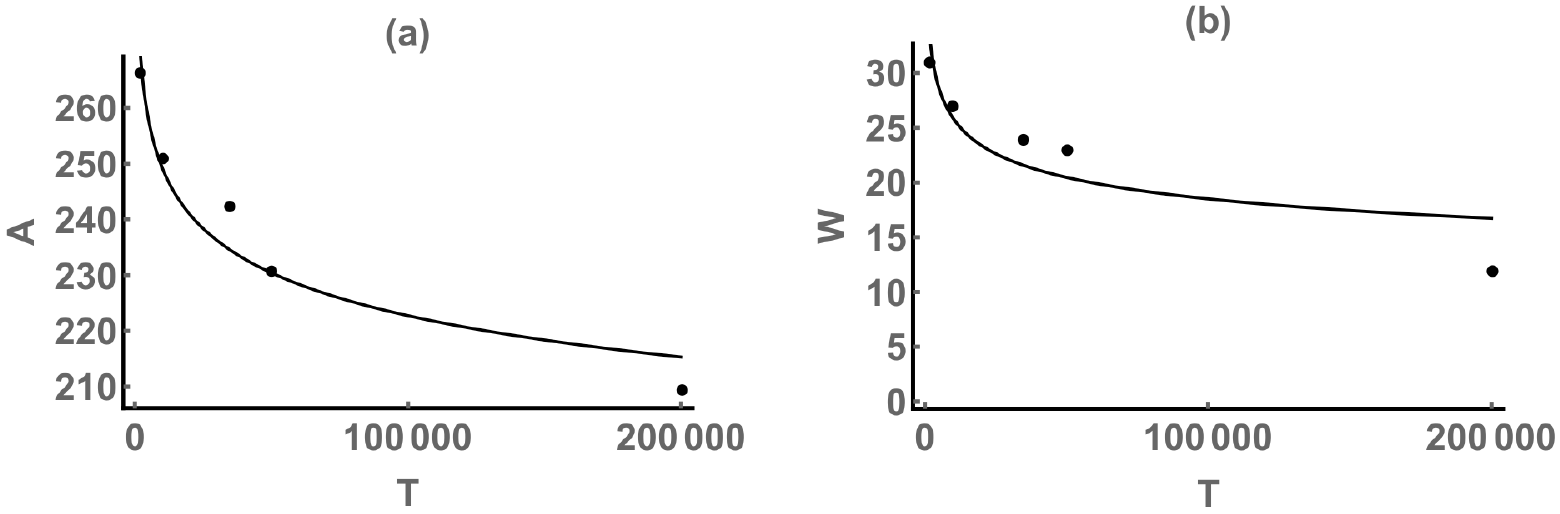}
\caption{
The island area (a) and the width of the island's flat top terrace (b) as functions of the dimensionless time, for a typical metastable decaying island on Pb(111) surface. 
Power-law fits are shown by the solid curves: (a) $A=391.174\ T^{-0.05}$,  (b) $W=98.584\ T^{-0.145}$.
}
\label{F10_1}
\end{figure}

In Fig. \ref{F10_1} we plot the time dependence of the area of the decaying island in Fig. \ref{F9}(e), as well as the width of the island's flat top terrace.
We considered the time interval during which the island keeps it characteristic electronic shape. 
The area scales as $A=391.174\ T^{-0.05}$, and the width as $W=98.584\ T^{-0.145}$. For a growing island, the approximate reciprocal dependencies hold.
Comparing to the data on Ag islands in Sec. \ref{Ag(110)}, it is apparent that Pb electronic islands grow or decay much slower. We again ascribe this difference to the large
lattice mismatch stress in Pb films; indeed, it has been reported that such stress has the effect of slowing the morphological evolution \cite{ABFFR}.

\begin{figure}[h]
\centering
\includegraphics[width=6.0in]{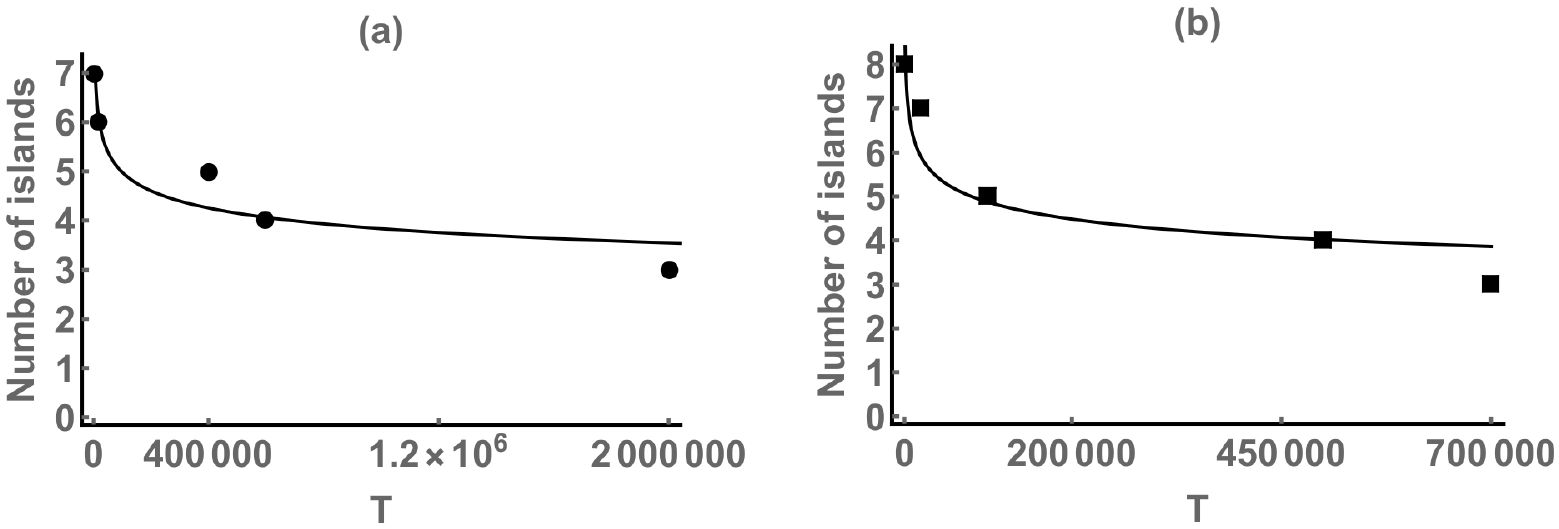}
\caption{
The number of flat-top and round-top islands, NI, on Pb(111) surface vs. the time. (a): Trial 12, (b): Trial 16.
Power-law fits are shown by the solid curves: (a) NI$=18.37\ T^{-0.113}$,  (b) NI$=19.24\ T^{-0.119}$. Data for the plot in (b) is taken from
\emph{Pb(111)Movie-Trial16}.
}
\label{F10}
\end{figure}

Last, Fig. \ref{F10} shows the number of electronic (flat-top) and non-electronic (rounded top) Pb islands vs. the time, for two representative trials. From the fits, 
the distance between the islands scales with time roughly as  $\langle L\rangle \approx T^{0.116}\sim T^{1/9 \div 1/8}$. Here the exponent is about two times smaller than
the one computed for coarsening of Ag islands.

%
%
%
%
%
%
%


In summary, this study provides the extensive computational results that show how the film morphology transitions between the metastable states of different ``magic" heights and
the routes by which an island reaches its final stable height, point to the co-existence of 3D and 2D islands, and quantify the time dependence of the island density and the 
island area during the collective evolution of quantum nanoislands on the surfaces of an ultrathin metal films.
The model can be extended to a 3D setup, material deposition, and the interfacially engineered systems \cite{RMC}.

\vspace{0.5cm}
\noindent
{\bf Supplementary Material}

\noindent
Movies \emph{Ag(110)Movie-Trial1.avi}, \emph{Ag(110)Movie-Trial3.avi}, \emph{Ag(110)Movie-Trial9.avi}, and \emph{Pb(111)Movie-Trial16.avi} show the computed evolution of the surface morphology
from the initial condition, see Fig. \ref{F2} for example, to the final state comprising of one or two quantum islands. 

\vspace{0.5cm}
\noindent
{\bf ACKNOWLEDGMENTS}

\noindent
MK's research was supported by Kentucky NSF EPSCoR (Grant No. 3200000271-18-069) and by WKU Research Foundation (QTAG grant).
DP's 
research was supported by Kentucky NSF EPSCoR (Grant No. 3200000271-18-069). VH's research was supported by the
Perm Region Ministry of Science and Education (Grant C-26/798).

\section{Appendix}

The model for the elastic strain energy in this paper is adapted from Ref. \cite{SDV}. The thin film (Pb or Ag) is considered epitaxially strained and dislocation-free, deposited on a 
rigid substrate. The assumption of a rigid substrate is appropriate for modeling the effects of the epitaxial stress on morphology evolution in these films, 
since the ratio of the shear moduli $\rho=\mu_F/\mu_S$ (F: film; S: substrate) is small \cite{SDV}.
For instance, $\rho=56\times 10^9$ erg/cm$^3$$/480\times 10^9$ erg/cm$^3$$=0.117$ for Pb/Cu, $\rho=56\times 10^9$ erg/cm$^3$$/641\times 10^9$ erg/cm$^3$$=0.087$
for Pb/Si, and $\rho=300\times 10^9$ erg/cm$^3$$/820\times 10^9$ erg/cm$^3$$=0.37$ for Ag/Fe.

To render the surface diffusion equation (\ref{MullinsEq}) dimensionless and thus reduce the number of parameters, we use the following scales: $x=LX$, $h=sH$, $t=\tau T/\alpha^4$, 
$\gamma=\gamma_{bf} \bar \gamma$, where $L=\gamma_{bf}\left(1-E_{33}^0\right)/4\mu_F\left(E_{33}^0\right)^2$ 
and $\tau=L^4/\mathcal{D}\gamma_{bf}$, with $E_{33}^0=\epsilon (1+\nu_F)/(1-\nu_F)$  \cite{SDV}, $\nu_F$ the Poisson's ratio, $\alpha=s/L$, 
and $\epsilon$ the misfit of the film.
Taking Ag film as the example, $\gamma_{bf}=515$ erg/cm$^2$ (see Fig. \ref{F1}(c)), $\nu_F=0.37$, $\mu_F=300\times 10^9$ erg/cm$^3$, and $\epsilon=1\%$ \cite{JP},
gives $L=8.9\times 10^{-7}$ cm. Thus $\alpha=1.95\times 10^{-8}$ cm$/8.9\times 10^{-7}$ cm$=0.022\ll 1$ and, as in Refs. \cite{SDV,Korzec,Korzec1,JEM_Khenner,MSMSE_Khenner}, 
we expand the adimensionalized form of Eq. (\ref{MullinsEq}) in $\alpha$ and then retain the dominant contributions. 
(From a mathematical standpoint, a multiplication by $\alpha$ emerges at each partial differentiation of $h(x,t)$ with respect to $x$, since 
$\partial h/\partial x=(s/L)\partial H/\partial X=\alpha H_X$.)
The expansion in the small parameter $\alpha$
is known as thin-film expansion, or approximation \cite{SDV,Korzec,Korzec1,ThinFilms}. The resultant evolution PDE reads
\begin{eqnarray}
H_T &=& -\bar \gamma(H) H_{XXXX} - 2A(H)H_X H_{XXX}-\left[A(H)H_{XX}+B(H)\left(1-H_X^2\right)\right]H_{XX} + C(H)H_X^2 \nonumber\\
&+& \mathcal{E} \left(HH_{XX}+\frac{H_X^2}{2}\right)_{XX} + \mathcal{E} E_{0004}\left(H^3H_{XXXX}\right)_{XX}+ \mathcal{E} E_{0013}\left(H^2H_XH_{XXX}\right)_{XX}.
\label{FinalPDE}
\end{eqnarray}
Here
\begin{equation}
\bar \gamma(H) = 1 + \frac{G_0}{(1+H)^2}\cos{\Omega_1 H}\cos{\Omega_2 H}-\frac{G_1}{1+H}
\label{gammaQSENdim}
\end{equation}
is the dimensionless surface energy (see Eq. (\ref{gammaQSE})). The parameters in Eq. (\ref{gammaQSENdim}) are $G_0=g_0/\gamma_{bf}$, $G_1=g_1/\gamma_{bf}$, $\Omega_1=s \omega_1$, 
and $\Omega_2=s \omega_2$. Also, $A(H)$, $B(H)$,
and $C(H)$ are the oscillatory functions of $H$ that contain the parameters $G_0$, $G_1$, $\Omega_1$, and $\Omega_2$. 
The terms in the first line of 
Eq. (\ref{FinalPDE}) stem from the chemical potential $\mu$ (Eq. (\ref{mu_kappa_mu_wet})), while the terms in the second line stem from the strain energy $\Sigma$.
The strain-related dimensionless parameters are
\begin{equation}
\mathcal{E}=\frac{4\left(E_{33}^0\right)^2}{\left(1-E_{33}^0\right)},\ E_{0013}=\frac{3+4\nu_F}{1-\nu_F},\ \mbox{and}\ E_{0004}=\frac{E_{0013}}{6}.
\label{ndimpara}
\end{equation}
We omitted three other nonlinear terms that are proportional to $\mathcal{E} E_{ijkl}$ \cite{SDV}, since for our
problem geometry and the parameters, their numerical contributions are negligible in comparison to the last two nonlinear terms in the second line of Eq. (\ref{FinalPDE}).
The only term among all $\mathcal{E} E_{ijkl}$ terms that contributes to the linear stability of the film is the retained $\mathcal{E} E_{0004}$ term.

The computational domain $R:X\in[0,X_{max}]$ for Eq. (\ref{FinalPDE}) and the random initial condition are parametrized as follows. Using again Ag film as example, 
we take the lateral film dimension 5 $\mu$m, which using $L\sim 10^{-6}$ cm (see the estimate above) gives $X_{max}=500$. The initial condition shown in Fig. \ref{F2} 
has on average 30 round-top islands per the domain $R$, which translates into a typical lateral dimension 170 nm of the island at its base. 
The tip of an island has the radius of curvature of several nanometers.
At the endpoints of the computational domain the periodic boundary conditions are imposed.
Eq. (\ref{FinalPDE}) is computed using the Method of Lines. Space discretization is done pseudospectrally, employing on average 3000 grid points on $R$. 
Integration in time of the system of the ordinary differential equations is done by the (stiff) BDF method.


\end{document}